\documentclass[twocolumn,showpacs,preprintnumbers,amsmath,amssymb,prb]{revtex4}
\usepackage[dvips]{color,graphics,epsfig,rotating}
\usepackage{graphicx}
\usepackage{dcolumn}
\usepackage{bm}

\begin{document}

\title{Positron potential and wavefunction in LaFeAsO}

\author{H. Takenaka}
\affiliation{Materials Science and Technology Division,
Oak Ridge National Laboratory, Oak Ridge, Tennessee 37831-6114} 

\author{D.J. Singh}
\affiliation{Materials Science and Technology Division,
Oak Ridge National Laboratory, Oak Ridge, Tennessee 37831-6114} 

\date{\today} 

\begin{abstract}
We report calculations of the positron potential and wavefunction
in LaFeAsO. These calculations show that the positron wavefunction
does sample the entire unit cell although it is largest in the
interstices of the La layer adjacent to As atoms. The implication
is that angular correlation of annihilation radiation (ACAR) is a viable
probe of the Fermi surfaces in this material. The results also apply
to positive muons, and indicate that these will be localized in
the La layer adjacent to As.
\end{abstract}


\maketitle

\section{introduction}

The discovery \cite{kamihara}
of high temperature superconductivity in a family
of layered oxypnictides, prototype LaFeAs(O,F) with
critical temperatures exceeded only by cuprates has stimulated
considerable interest both in determining the chemical dependence
of the properties and in understanding the mechanism for superconductivity.
Central to this discussion is the underlying electronic structure.

\begin{figure}
\includegraphics[width=3.2in,angle=0]{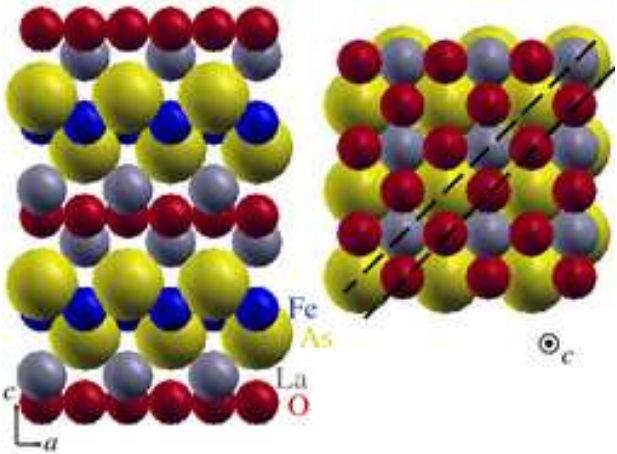}
\caption{\label{struct} (Color online)
Structure of LaFeAsO viewed along [100]
(left) and [001] (right) directions. The
dashed lines in the right panel show the two cuts
for which potentials and positron densities are plotted
in Figs. \ref{fig-vcoul}, \ref{fig-vpos}, and \ref{fig-pos}.
}
\end{figure}

These materials occur in a tetragonal structure, \cite{kamihara,quebe,struct}
based on a square lattice of Fe coordinated in such a way that the
unit cell is based on a c(2x2) doubling of the Fe square lattice
structure (see Fig. \ref{struct}).
First principles calculations done within density functional theory
predict five small sheets of Fermi surface in the undoped compound,
LaFeAsO: two 2D electron cylinders at the zone corner ($M$),
two heavier 2D hole cylinders at the zone center ($\Gamma$) and
a still heavier 3D hole pocket, which intersects the hole cylinders.
\cite{singh-du}
Because of the heavy masses of the bands, especially the hole bands,
the density of states is high even though the carrier density is low.

Superconductivity generally requires a pairing interaction.
Direct calculations of the electron-phonon interaction have shown that
it is modest (the coupling is 
$\lambda \sim$0.2)
and in particular is far
too weak to explain the observed critical temperatures.
\cite{boeri,mazin}
In unconventional superconductors the
{\bf k} dependence of the pairing
interaction plays a central role as this
{\bf k} dependence acting on the Fermi surface
determines the symmetry of the superconducting state that emerges.
The 2D cylinders are roughly nested and as such may be expected to
lead to {\bf k} dependence of properties. In fact a spin density
wave associated with the nesting has been predicted and
observed in the undoped compound but not so far in the doped superconducting
material.
\cite{mazin,dong,cruz,zhu}
Models of superconductivity associated with spin fluctuations
deriving from this nesting have been discussed. \cite{mazin}
However, small Fermi surfaces, which are derived from states near
band edges, are in general particularly sensitive to details of the crystal
structure and are also more sensitive to disorder and perhaps to
many body effects.
As such, it is important to determine the Fermi surface from experiment.

The most common experimental probes of Fermi surfaces in metals are
(1) quantum oscillation measurements, such as de Haas van Alphen and
Shubnikov de Haas,
(2) angle resolved photoelectron spectroscopies (ARPES),
(3) positron annihilation, in particular angular correlation of positron
annihilation radiation (ACAR)
and (4) Compton scattering.
These techniques are complimentary.
Quantum oscillation techniques are the methods of choice since
they are direct and have the highest resolution when they 
are practical, but they require very high quality samples with long
mean free paths. ARPES is particularly applicable to 2D materials,
where it gives a direct map of the Fermi surface and does not require
such high mean free paths. It is however surface sensitive and requires
that clean unreconstructed surfaces characteristic of bulk can be made.
Positron ACAR and Compton scattering are bulk techniques with lower resolution
than ARPES or quantum oscillations, but with much less stringent sample
quality issues. ACAR is the more common of these two techniques
and may be the most readily applicable method if crystals
of the LaFeAs(O,F) phases become available.
However, the sensitivity of ACAR measurements depends on the overlap
of the positron wavefunction with the electronic states at the Fermi
level. In these materials those states are primarily Fe $d$ states
modestly hybridized with As $p$ states.
A related issue was recognized in high $T_c$ cuprate superconductors, 
\cite{von-stetten,singh-89}
where it was found that the sensitivity of ACAR to CuO$_2$ planes
depended on the particular material, and in particular that ACAR was
not sensitive to the CuO$_2$
derived electronic states in YBa$_2$Cu$_3$O$_7$.
The purpose of the present paper is to report positron wavefunctions
for LaFeAsO.

\begin{figure}
\includegraphics[width=3.2in,angle=0]{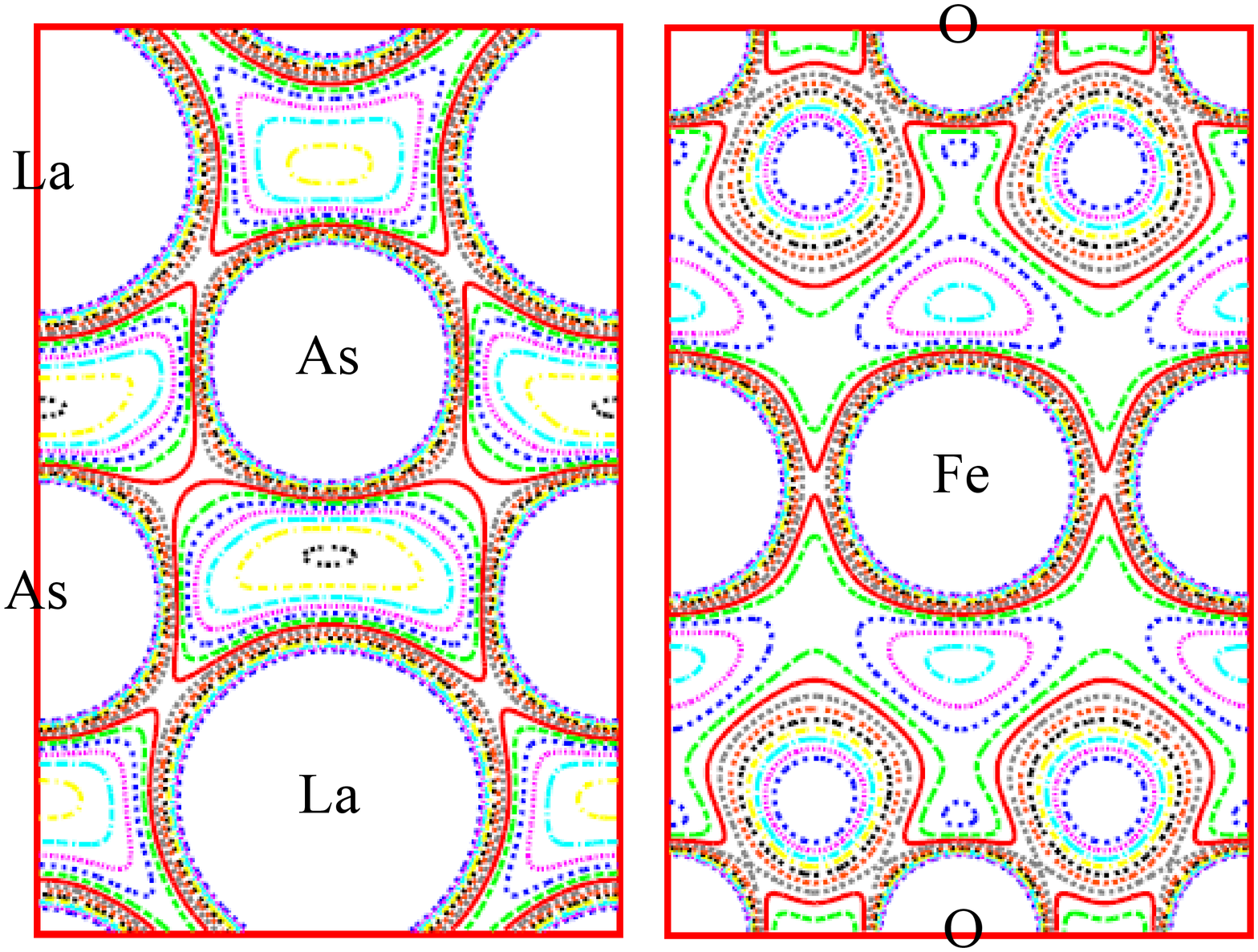}
\caption{\label{fig-vcoul} (Color online)
Inverted Coulomb potential of LaFeAsO shown in (110) planes
(see Fig. \ref{struct}). The potential is divergently repulsive
inside the atom cores. The contours shown span a range of 5.65 eV and
are equally spaced.
}
\end{figure}

\section{approach}

The present calculations were done using the general potential
linearized augmented planewave method including
local orbitals. \cite{singh-book,singh-lo}
The self consistent electronic structure was first calculated using
the experimental lattice parameters but the LDA relaxed internal 
coordinates for La and As.
Computational parameters for this
were set as described in Ref. \onlinecite{singh-du}.
The positron calculation was then done using the potential
generated from the electronic charge density. In particular,
the bulk positron wavefunction was computed in the inverted
Coulomb potential to which a correlation term was added,

\begin{equation}
\label{vpositron}
V^+[n] = -V_{Coul}[n] + V_{corr}[n] ,
\end{equation}

\noindent
where $V^+$ is the potential to be used for the positron wavefunction
calculations, $V_{Coul}$ is the Coulomb potential from
the electronic calculation
and $V_{corr}$ is an electron positron correlation function. All of these are
functionals of the electron density, but not the positron density, since
for the bulk positron wavefunction a single positron is distributed over
a macroscopic sample, and therefore has vanishing density.
The correlation function, $V_{corr}$ is an attractive potential that
arises from the response of the electrons to the positron.
While the sum rule that the correlation peak, which is the analogy of the
exchange-correlation hole of the electron gas, has unit charge, the
magnitude of this buildup is not bounded, while the electron-electron
hole is bounded by the local density. Therefore, the electron-positron
correlation potential is thought to be less well behaved than the
electron-electron correlation. Nonetheless, local density parameterizations
have been developed. Here we use the parameterization of Boronski and
Nieminen, \cite{boronski} applied as in Ref. \onlinecite{takenaka}.
We note that a similar approach can in principle be applied to other
charged particles, in particular positive muons, $\mu^+$. In that case
the potential for the $\mu^+$ can be formally written as for the
positron, except that because of the higher mass of the $\mu^+$ the
reduced mass of an $\mu^+$ - $e^-$ pair is higher, and therefore the
correlation potential will be stronger and more difficult to reliably
parameterize. \cite{karlsson,manninen,nieminen-mu}

\section{results and discussion}

\begin{figure}
\includegraphics[width=3.2in,angle=0]{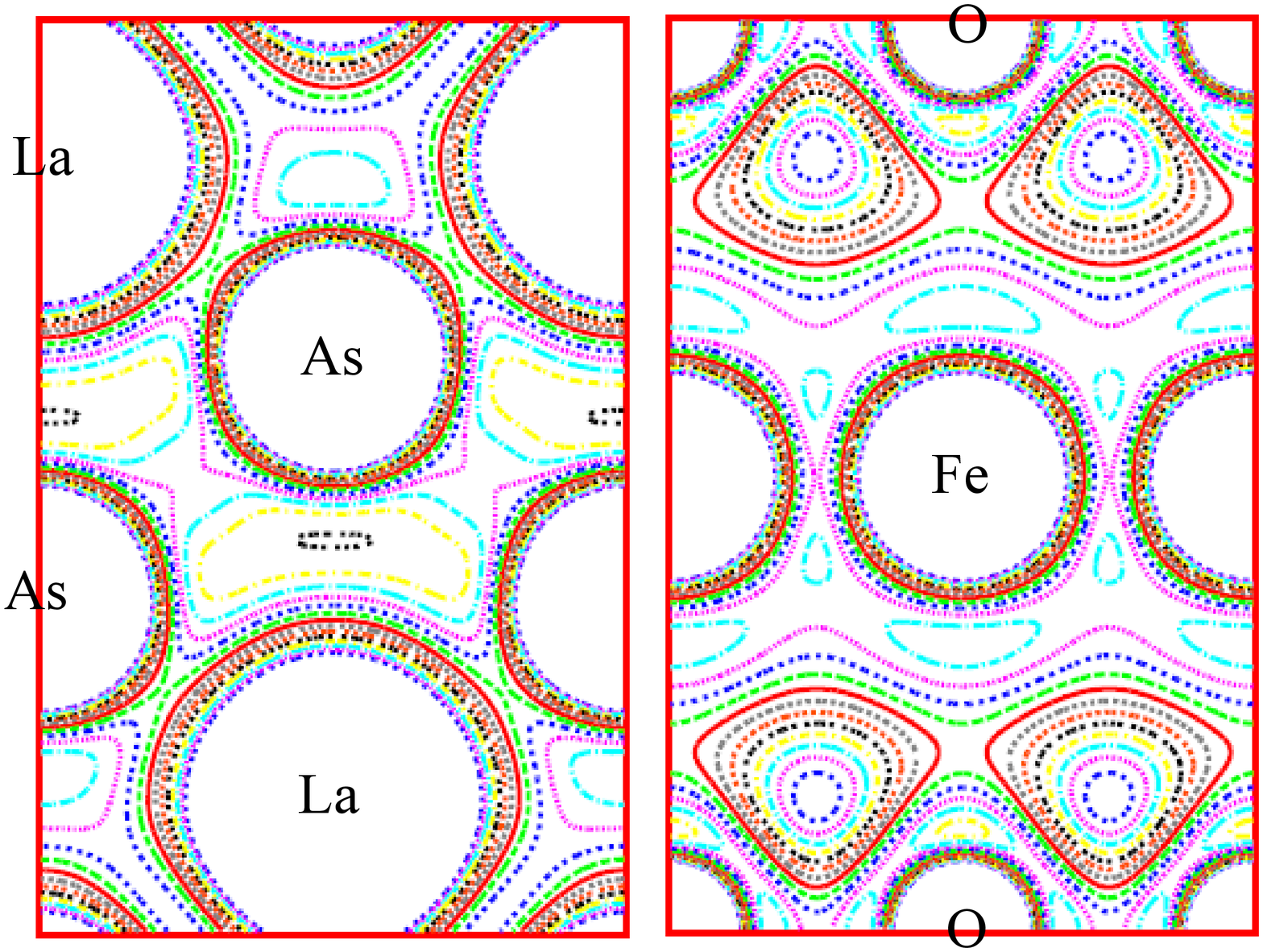}
\caption{\label{fig-vpos} (Color online)
Positron potential of LaFeAsO
(see Fig. \ref{struct}).
The potential is divergently repulsive inside the atom cores. The
contours span a range of 5.92 eV and are equally spaced.
}
\end{figure}

\begin{figure}
\includegraphics[width=3.2in,angle=0]{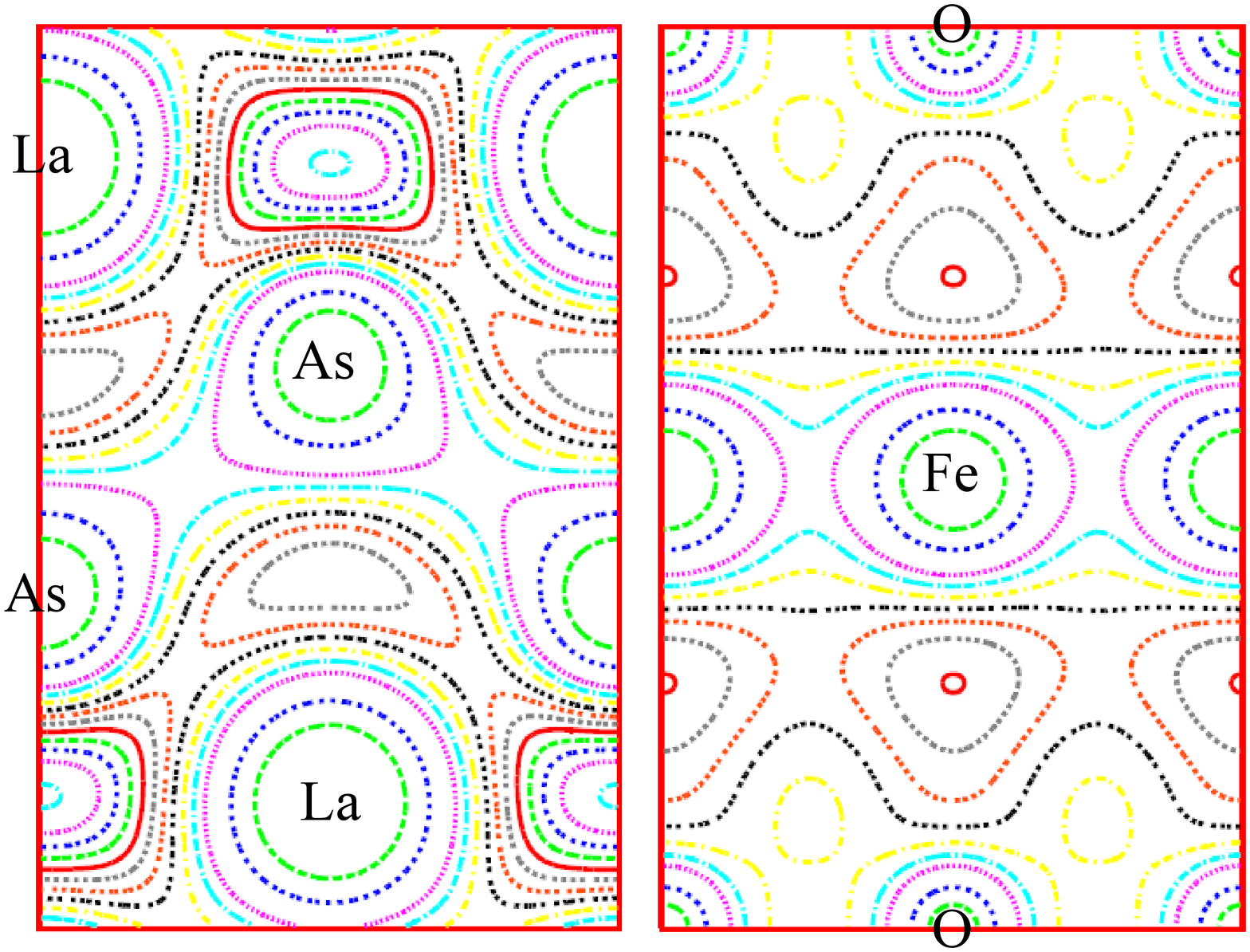}
\caption{\label{fig-pos} (Color online)
Positron density of LaFeAsO
(see Fig. \ref{struct}). The positron density goes to zero
in the atom cores. The contours are equally spaced and range
from 0.00008 to 0.00265 in units of positrons per cubic Bohr,
normalized to one positron per unit cell.
}
\end{figure}

The main results of this work are shown in Figs. \ref{fig-vcoul},
\ref{fig-vpos} and \ref{fig-pos}. These show respectively
the inverted Coulomb potential,
the positron potential, $V^+$ and the positron density, which is the
square of the positron wavefunction, normalized to one positron per cell
for convenience.

As may be seen, the inverted Coulomb potential is most attractive in
the interstices of the Fe-As part of the unit cell.
A positron localized there would be most sensitive to the
electronic states near the Fermi energy. However, the electron
density is higher on the other side of the As ions, in the interstices of the
the La layer.
Because of this, the addition of the correlation term favors that site.
The absolute minimum of the potential remains near the Fe when the correlation
potential is added.
However, the position in the interstices of the La
layer above As is larger, and the additional attraction provided by
the correlation function is enough to shift the maximum of the positron
density to that position.
Still, this is not strong enough
to fully pull the positron density away from the Fe planes.
Thus as
may be seen in Fig. \ref{fig-pos}, the positron density is highest in the
interstices of the
La plane nearest to As following the potential, $V^+$.
The calculated positron lifetime, obtained as in Ref. \onlinecite{takenaka},
is 163 ps.

This result means that positrons will mainly sample the electronic states
away from Fe. However, this is not as strong a localization as was found
in YBa$_2$Cu$_3$O$_7$ (Ref. \onlinecite{singh-89}), and in particular some
significant density may still be seen around the Fe. Furthermore, it should
be noted that the states near the Fermi level have some hybridization
between Fe and As, similar to an oxide electronic structure. As such, the
fact that the positrons are sensitive to As will give them some
sensitivity to the Fermi surface. Thus is would seem that ACAR is a
viable technique for detecting the Fermi surfaces of LaFeAs(O,F).

Turning to positive muons,
there have been several recent muon spin rotation studies
of these materials. \cite{drew,khasanov,carlo}
These studies have yielded quite useful insights into the magnetism
of the materials, showing signatures both of the spin density wave
and of rare earth magnetism and in addition have been useful in
establishing the penetration depth.
As mentioned the correlation potential for
muons will be more strongly attractive than for positrons.
Furthermore, because of their heavier mass, muons will not be
delocalized in the lattice, but rather will localize, similar to a 
proton. This will lead to a change in the electron density, which
should be treated self-consistently. Qualitatively however this
effect will also amount to an increased tendency for the muon to be
located where the electron density is high. This means that the effect
that for the positron adding $V_{corr}$ to the inverted Coulomb potential
draws the positron away from the Fe layer will be enhanced for positive
muons so that they will be drawn away even more strongly.
Thus we may conclude that positive muons probe the interstices in the La
layer adjacent to As, and therefore that are most sensitive to the rare
earth site.

\section{summary}
Density functional calculations of the positron wavefunction in
LaFeAsO show that positrons probe the entire unit cell, but are
mainly located in the interstices of the La layer adjacent to As
in this structure. Considering that there is some overlap with
the Fe layer and considering that the Fe $d$ derived
electronic states at the Fermi
energy are hybridized with As, albeit modestly,
we conclude that positron ACAR can be used to measure Fermi surfaces
in this material.

\acknowledgements

We are grateful for helpful discussions with D.G. Mandrus and B.C. Sales.
This work was supported by the Department of Energy, Division of
Materials Sciences and Engineering.


\begin{references}
\bibitem{struct}
At room temperature the structure is tetragonal, but in the
undoped compound there is a distortion to a lower symmetry
monoclinic or orthorhombic structure as the temperature is lowered;
see ref. \onlinecite{cruz}.
\end{references}
\end{document}